\documentclass[twocolumn]{aastex631}
\usepackage[utf8]{inputenc}
\pdfoutput=1 
\usepackage{amsmath,amstext}
\usepackage[T1]{fontenc}

\newcommand{\hii}         {\mbox{\rm \ion{H}{2}}}
\newcommand{\htwo}        {\mbox{H$_{2}$}}
\newcommand{\jone}        {\mbox{$J=1-0$}}

\newcommand{\kmpers}      {\mbox{\rm km~s$^{-1}$}}

\newcommand{\percmcu}     {\mbox{\rm cm$^{-3}$}}
\newcommand{\msun}        {\mbox{\rm M$_\odot$}}

\newcommand{\msunperyr}   {\mbox{\rm M$_\odot$~yr$^{-1}$}}

\newcommand{\xco}         {\mbox{$X_{\rm CO}$}}

\newcommand{\xcounits}    {\mbox{\rm cm$^{-2}$(K km s$^{-1}$)$^{-1}$}}

\newcommand{\Kkmpers}     {\mbox{\rm K km s$^{-1}$}}

\newcommand{\percmsq}     {\mbox{cm$^{-2}$}}

\newcommand{\rev}[1]{{#1}}

\usepackage{graphicx} 

\begin{document}

\newcommand{\AAPF}{\altaffiliation{NSF Astronomy and Astrophysics Postdoctoral Fellow}}

\newcommand{\Arizona}{\affiliation{Steward Observatory, University of Arizona, Tucson, AZ 85721, USA}}

\newcommand{\Maryland}{\affiliation{Department of Astronomy, University of Maryland, College Park, MD 20742, USA}}

\newcommand{\JSI}{\affiliation{Joint Space-Science Institute, University of Maryland, College Park, MD 20742, USA}}

\newcommand{\OSU}{\affiliation{Department of Astronomy, The Ohio State University, Columbus, OH 43210, USA}}

\newcommand{\OSUphysics}{\affiliation{Department of Physics, The Ohio State University, Columbus, OH 43210, USA}}

\newcommand{\OSUCCAPP}{\affiliation{Center for Cosmology and Astro-Particle Physics, The Ohio State University, Columbus, OH 43210, USA}}

\newcommand{\NRAO}{\affiliation{National Radio Astronomy Observatory, 520 Edgemont Road, Charlottesville, VA 22903}}

\newcommand{\NRAOSocorro}{\affiliation{National Radio Astronomy Observatory, P.O. Box O, 1003 Lopezville Road, Socorro, NM 87801, USA}}

\newcommand{\Jansky}{\altaffiliation{Jansky Fellow of the National Radio Astronomy Observatory}}

\newcommand{\Concepcion}{\affiliation{Departamento de Astronom\'ia, Universidad de Concepci\'on, Barrio Universitario, Concepci\'on, Chile}}

\newcommand{\Kansas}{\affiliation{Department of Physics and Astronomy, University of Kansas, 1251 Wescoe Hall Dr., Lawrence, KS 66045, USA}}

\newcommand{\MPIA}{\affiliation{Max-Planck-Institut f\"ur Astronomie, K\"onigstuhl 17, 69120 Heidelberg, Germany}}

\newcommand{\IPAC}{\affiliation{IPAC, California Institute of Technology, 1200 East California Boulevard, Pasadena, CA 91125, USA}}

\newcommand{\STScI}{\affiliation{Space Telescope Science Institute, 3700 San Martin Drive, Baltimore, MD 21218, USA}}

\newcommand{\Swinburne}{\affiliation{Centre for Astrophysics and Supercomputing, Swinburne University of Technology, Hawthorn, VIC 3122, Australia}}

\newcommand{\ASTROTD}{\affiliation{ARC Centre of Excellence for All Sky Astrophysics in 3 Dimensions (ASTRO 3D)}}

\newcommand{\ITA}{\affiliation{Universit\"{a}t Heidelberg, Zentrum f\"{u}r Astronomie, Institut f\"{u}r Theoretische Astrophysik, Albert-Ueberle-Str. 2, D-69120 Heidelberg, Germany}}

\newcommand{\IWR}{\affiliation{Universit\"{a}t Heidelberg, Interdisziplin\"{a}res Zentrum f\"{u}r Wissenschaftliches Rechnen, Im Neuenheimer Feld 205, D-69120 Heidelberg, Germany}}

\newcommand{\UTA}{\affiliation{Department of Astronomy, The University of Texas at Austin, 2515 Speedway, Stop C1400, Austin, TX 78712, USA}}

\newcommand{\Princeton}
{\affiliation{Department of Astrophysical Sciences, Princeton University, Princeton, NJ 08544, USA}}

\newcommand{\UWYO}
{\affiliation{Department of Physics \& Astronomy, University of Wyoming, Laramie, WY 82071, USA}}

\newcommand{\SOFIA}
{\affiliation{Stratospheric Observatory for Infrared Astronomy, NASA Ames Research Center, Mail Stop 204-14, Moffett Field, CA 94035, USA}}

\newcommand{\JPL}
{\affiliation{Jet Propulsion Laboratory, California Institute of Technology, 4800 Oak Grove Dr., Pasadena, CA 91109, USA}}

\newcommand{\Leiden}
{\affiliation{Leiden Observatory, Leiden University, P.O.~Box 9513, 2300~RA~Leiden, The Netherlands}}

\newcommand{\UCSD}
{\affiliation{Department of Astronomy \& Astrophysics, University of California, San Diego, La Jolla, CA 92093, USA}}

\newcommand{\UMN}
{\affiliation{Minnesota Institute for Astrophysics, 
University of Minnesota,
Minneapolis, MN 55455}}

\newcommand{\UdeC}{\affiliation{Departamento de Astronom\'ia, Universidad de Concepci\'on, Barrio Universitario, Concepci\'on, Chile}}

\newcommand{\KU}{\affiliation{Department of Physics and Astronomy, University of Kansas, 1251 Wescoe Hall Dr., Lawrence, KS 66045, USA}}

\newcommand{\ESOST}{\affiliation{European Space Agency, c/o STScI, 3700 San Martin Drive, Baltimore, MD 21218, USA}}

\newcommand{\INAOE}{\affiliation{Instituto Nacional de Astrof\'{\i}sica, \'Optica y Electr\'onica, Luis Enrique Erro 1, Tonantzintla 72840, Puebla, Mexico}}

\newcommand{\NMMT}{\affiliation{New Mexico Institute of Mining and Technology, 801 Leroy Place, Socorro, NM 87801, USA}}

\newcommand{\Gent}{\affiliation{Sterrenkundig Observatorium, Ghent University, Krijgslaan 281 - S9, B-9000 Gent, Belgium}}

\newcommand{\UToledo}{\affiliation{Ritter Astrophysical Research Center, University of Toledo, Toledo, OH 43606, USA}}

\newcommand{\UGR}{\affiliation{Dept. F\'isica Te\'orica y del Cosmos, Universidad de Granada, 18071, Granada, Spain }}

\newcommand{\OAN}{\affiliation{Observatorio Astronómico Nacional (IGN), C/Alfonso XII, 3, E-28014 Madrid, Spain}}

\newcommand{\YS}{\affiliation{Centro de Desarrollos Tecnológicos, Observatorio de Yebes (IGN), 19141 Yebes, Guadalajara, Spain}}

\title{JWST Observations of Starbursts: Polycyclic Aromatic Hydrocarbon Emission at the Base of the M~82 Galactic Wind}

\author[0000-0002-5480-5686]{Alberto D. Bolatto}
\Maryland
\JSI
\author[0000-0003-2508-2586]{Rebecca C. Levy}
\AAPF
\Arizona

\author[0000-0003-1356-1096]{Elizabeth Tarantino}
\STScI

\author[0000-0003-4850-9589]{Martha L. Boyer}
\STScI

\author[0000-0003-0645-5260]{Deanne B. Fisher}
\Swinburne
\ASTROTD

\author[0000-0002-9511-1330]{Serena A. Cronin}
\Maryland

\author[0000-0002-2545-1700]{Adam K. Leroy}
\OSU

\author[0000-0002-0560-3172]{Ralf S. Klessen}
\ITA
\IWR

\author[0000-0003-1545-5078]{J.D. Smith}
\UToledo

\author[0000-0002-4153-053X]{Danielle A. Berg}
\UTA

\author[0000-0002-5666-7782]{Torsten B\"oker}
\ESOST

\author[0000-0002-3952-8588]{Leindert A. Boogaard}
\MPIA

\author[0000-0002-0509-9113]{Eve C. Ostriker}
\Princeton

\author[0000-0003-2377-9574]{Todd A. Thompson}
\OSU
\OSUCCAPP
\OSUphysics

\author[0000-0001-8224-1956]{Juergen Ott}
\NRAOSocorro

\author[0000-0003-4023-8657]{Laura Lenki\'{c}}
\SOFIA
\JPL

\author[0000-0002-1790-3148]{Laura A. Lopez}
\OSU
\OSUCCAPP

\author[0000-0002-1790-3148]{Daniel~A.~Dale}
\UWYO

\author[0000-0002-3158-6820]{Sylvain Veilleux}
\Maryland
\JSI

\author[0000-0001-5434-5942]{Paul P.~van der Werf}
\Leiden

\author[0000-0001-6708-1317]{Simon C.~O.~Glover}
\ITA

\author[0000-0002-4378-8534]{Karin M. Sandstrom}
\UCSD

\author[0000-0003-0605-8732]{Evan D.\ Skillman}
\UMN

\author[0000-0002-0302-2577]{John Chisholm}
\UTA

\author[0000-0002-5877-379X]{Vicente Villanueva}
\Maryland
\UdeC

\author[0000-0001-8490-6632]{Thomas S.-Y. Lai}
\IPAC

\author[0000-0002-2644-0077]{Sebastian Lopez}
\OSU
\OSUCCAPP

\author[0000-0001-8782-1992]{Elisabeth A.C. Mills}
\KU

\author[0000-0001-6527-6954]{Kimberly L.~Emig}
\Jansky
\NRAO

\author[0000-0003-3498-2973]{Lee Armus}
\IPAC

\author[0000-0002-4677-0516]{Divakara Mayya}
\INAOE

\author[0000-0001-9436-9471]{David S. Meier}
\NMMT
\NRAOSocorro

\author[0000-0001-9419-6355]{Ilse De Looze}
\Gent

\author[0000-0002-2775-0595]{Rodrigo Herrera-Camus}
\UdeC

\author[0000-0003-4793-7880]{Fabian Walter}
\MPIA

\author[0000-0003-1682-1148]{M\'onica Rela\~no}
\UGR

\author[0009-0001-5949-1524]{Hannah B. Koziol}
\UCSD

\author[0000-0003-1111-8066]{Joshua Marvil}
\NRAOSocorro

\author[0000-0002-9165-8080]{Mar\'ia J. Jim\'enez-Donaire}
\OAN
\YS

\author[0000-0002-4279-4182]{Paul Martini}
\OSU
\OSUCCAPP

\correspondingauthor{Alberto D. Bolatto}
\email{bolatto@umd.edu}

\begin{abstract}
We present new observations of the central 1~kpc of the M~82 starburst obtained with the {\em James Webb Space Telescope} (JWST) near-infrared camera (NIRCam) instrument at a resolution $\theta\sim0.05\arcsec-0.1\arcsec$ ($\sim1-2$~pc). The data comprises images in three mostly continuum filters (F140M, F250M, and F360M), and filters that contain [\ion{Fe}{2}] (F164N), \htwo\ $v=1\rightarrow0$ (F212N), and the 3.3 $\mu$m PAH feature (F335M). We find prominent plumes of PAH emission extending outward from the central starburst region, together with a network of complex filamentary substructure and edge-brightened bubble-like features. The structure of the PAH emission closely resembles that of the ionized gas, as revealed in Paschen $\alpha$ and free-free radio emission. We discuss the origin of the structure, and suggest the PAHs are embedded in a combination of neutral, molecular, and photoionized gas.
\end{abstract}


\shorttitle{JWST Observations of M~82 }

\shortauthors{Bolatto et al.}

\section{Introduction}
Local starburst galaxies are characterized by fast star formation causing rapid consumption of their gas reservoirs on typical time-scales of $\sim100$\,Myr. This `high activity' mode resembles the situation found at cosmic noon ($z\sim2$), where gas-rich, strongly star-forming galaxies dominate the cosmic star formation budget. Therefore it is important to understand the physics that drive this mode of star formation, something best done in nearby galaxies which can be studied in detail. Differences in the physical conditions of the interstellar medium (ISM) during starburst phases have been invoked as drivers for variations in the stellar initial mass function \citep[IMF, e.g.,][]{Conroy2012}, while concentrated star formation events can also drive multi-phase winds that drastically affect the evolution of a galaxy \citep{Heckman1990,Veilleux2005,Veilleux2020}, limiting gas available to fuel super-massive black holes and grow galaxies, and playing a key role in determining the galaxy mass function.

The higher rate of star formation per unit molecular gas mass observed in starbursts strongly suggests actual physical changes in the process of forming stars, not just a one-to-one scaling of the activity with the gas reservoir.  Starbursts occur due to the concentration of large masses of (molecular) gas, that when combined with shorter free-fall timescales, produce high star formation rates \citep[e.g.,][]{Krumholz2012,Leroy2015a}. High gas surface densities in starbursts imply that very high gas consumption rates are required for star formation feedback to prevent the collapse of their gas disks \citep{Ostriker2011}. 

These conditions alter how star formation happens, causing the fraction of stars formed in clusters to likely increase with the surface density of star formation \citep[e.g.,][]{Kruijssen2012,Krumholz2019}. Vigorous starbursts produce super star clusters \citep[SSCs, e.g.,][]{Herrera2012,Whitmore2014,Turner2015,Leroy2018,Emig2020}; massive (M$_\star \sim10^5$\,M$_\odot$), and compact ($R\sim1$\,pc) concentrations of stars that are the younger cousins to globular clusters \citep{Whitmore2005,PortegiesZwart2010,Linden2021}. The formation of SSCs is a particularly efficient way of converting gas into stars \citep{Krumholz2019}, likely characteristic of starburst environments. 

This highly concentrated star formation activity also results in massive ejections of material from the galaxy into the circumgalactic medium, which eventually quenches the starburst and curtails the growth of the galaxy \citep{Heckman1990}. Galactic winds are thought to regulate the growth of galaxies and thus shape the galaxy mass function, in part because they eject cool gas into the circumgalactic medium that is then no longer immediately available for star formation (although it may be reaccreted on long time scales), in part by preventing the accretion of cosmic web gas into the galaxy halo \citep[e.g.,][]{Mitchell2022}. The concentrated starburst and the launching of a wind are intimately related, and this may leave an imprint in their respective structures. For example, starburst rings may produce fast biconical hot outflows collimated along the ring axis sheathed in cooler material \citep{Nguyen2022}.
The mass loss rate of the wind --- key to understand its effect on galaxy evolution ---, is highly uncertain because it requires measuring the contribution of its different phases, specially the cooler ($T\lesssim10^4$~K) phases thought to carry the bulk of the mass. 

The key observations we present in this study concern the imaging of the base of the M~82 wind in polycyclic aromatic hydrocarbon (PAH) emission. PAHs are complex molecules and/or very small dust grains that pervade the cool ISM, absorbing UV and optical photons and re-emitting their energy in relatively broad spectral features (bands) throughout the mid-infrared spectrum. They play a key role in the heating of the neutral gas and possibly in the formation of H$_2$.  The 3.3\,$\mu$m PAH band is thought to be mostly due to small, neutral PAHs \citep{Draine2001,Maragkoudakis2020,Draine2021}. PAHs are extraordinarily resilient, being observed in close proximity to active galactic nuclei \citep[e.g.,][]{Garcia-Bernete2022}. They
are destroyed by shocks \citep{Micelotta2010a} or X-rays associated with luminous AGN \citep{Xie2022,Lai2023} and T-Tauri stars \citep{Siebenmorgen2010}, however, and 
are rapidly destroyed in hot gas \citep[e.g.,][]{Micelotta2010b}, therefore they trace the cooler phases of the ISM and their spectrum provides key information on the physical conditions in those phases. 

M~82 is part of the M~81 group located $3.6$\,Mpc away \citep{Freedman1994,Dalcanton2009}, and represents the prototypical example of an interaction-driven global starburst \citep{Yun1994,Mayya2006,deBlok2018} in a dwarf galaxy \citep[the mass of M~82 is approximately 10$^{10}$\,\msun,][]{Greco2012}, affording a uniquely detailed view of the astrophysical processes at the core of the starburst phenomenon. Modeling of near to mid-infrared spatially resolved spectroscopy suggests that M~82 experienced a first starburst episode in its central 500~pc between $8-15$ Myr ago peaking at 160~\msunperyr, followed by a second, somewhat more compact burst occurring in a circumnuclear ring and stellar bar $4-6$ Myr ago and peaking at 40~\msunperyr\ \citep{Colbert1999,ForsterSchreiber2003}. The far-infrared luminosity of $4\times10^{10}$~L$_\odot$ \citep{Herrera-Camus2018} corresponds to a current star formation rate of SFR~$\simeq12$~\msunperyr, broadly consistent with other recent estimates \citep{ForsterSchreiber2003}. 

M~82 not only hosts a large number of compact, massive clusters that represent a significant fraction of the star formation activity \citep{O'Connell1995,McCrady2003,Melo2005,Smith2006,McCrady2007,Mayya2008}, but it also has a powerful multiphase wind. This wind is visible in X-rays \citep{Strickland1997,Lopez2020}, hydrogen recombination \citep{Shopbell1998,Lokhorst2022}, far-infrared dust and line emission \citep{Engelbracht2006,Contursi2013,Beirao2015,Levy2023}, warm molecular hydrogen \citep{Veilleux2009,Beirao2015}, and cold molecular and neutral atomic gas \citep{Walter2002,Leroy2015b,Martini2018,Krieger2021}. The wind geometry corresponds to a truncated bi-cone with its southern lobe approaching us, and an opening angle of $\Omega_{w}\sim0.8\pi$\,sr \citep{Xu2023}. The main ionization mechanism of the wind is photoionization close to the starburst, with increasing shock ionization $1-2$~kpc away from it \citep{Shopbell1998}. The wind extends at least 11~kpc away from the galaxy \citep{Devine1999,Lehnert1999}, and it may reach out to 40~kpc \citep{Lokhorst2022}. Polarization work shows the existence of a large reflection nebula in H$\alpha$ \citep{Scarrott1991}, with a polarization fraction that increases to 30\% outwards, at 4~kpc from the nucleus \citep{Yoshida2019}. Kinematic analysis suggests that some of the dust causing the reflection 4~kpc away from the galaxy center is moving at velocities of $300-450$~\kmpers, faster than the molecular gas velocities observed closer in \citep{Leroy2015b}. The radiation reflected appears to have been emitted by a combination of two sources; one is the nucleus of M~82 as identified at 2.2\,$\mu$m (which dominates at large distances), the other coincident with the 3\,mm continuum peak west of the nucleus \citep{Yoshida2019}.

\begin{figure*}[t]
\centering
\includegraphics[width=\textwidth]{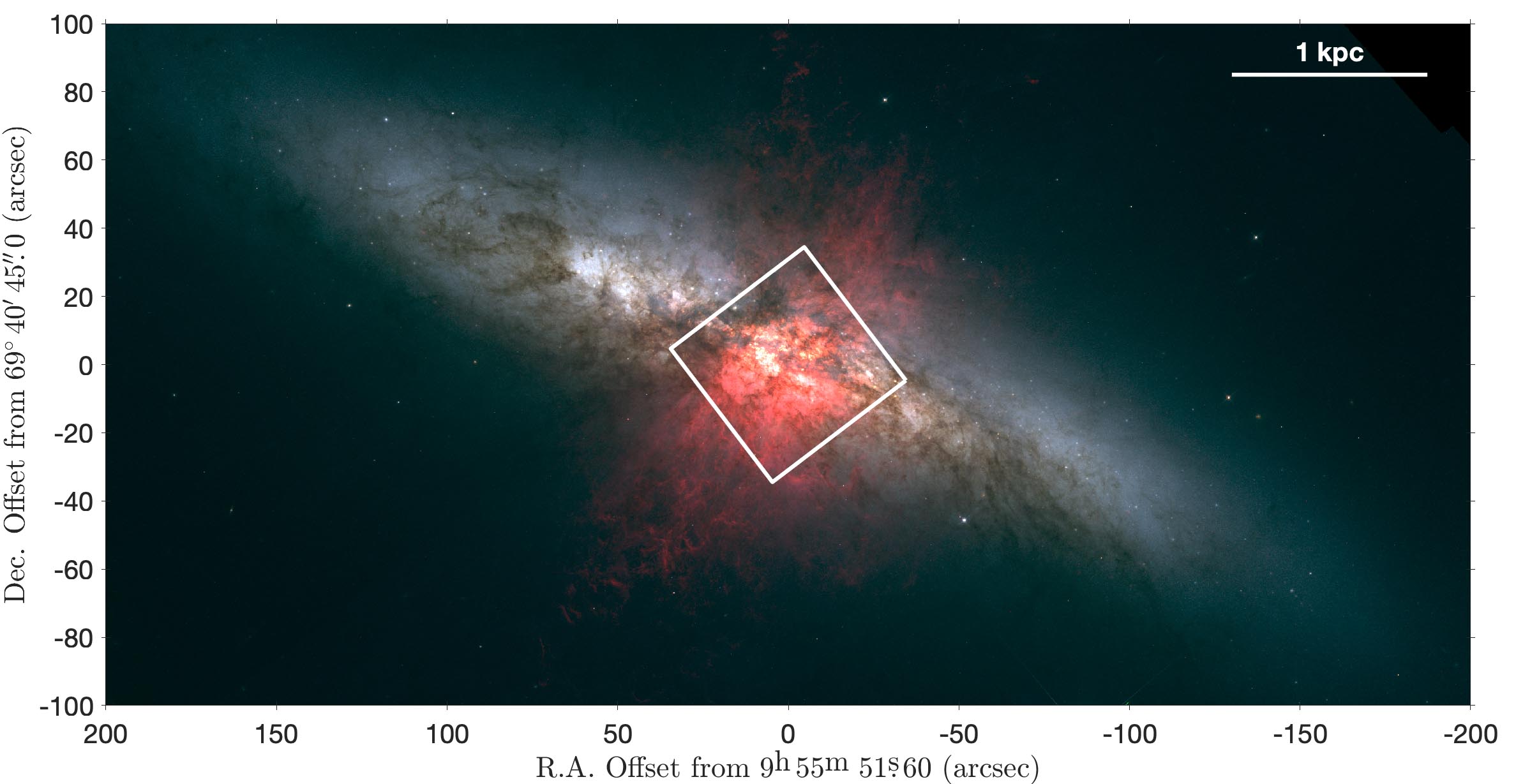}
\caption{Advanced Camera for Surveys (ACS) image of M~82 \citep{Mutchler2007} with the JWST NIRCam central mosaic footprint overlaid for reference. We use F658N (H$\alpha$), F555W, and F435W for red, green, and blue respectively. These  JWST observations cover the central region of the starburst and the base of the wind bi-cone. They are part of a larger mosaic imaging the wind and the starburst, which will be acquired in 2024.\label{fig:location}}  
\end{figure*}

In this manuscript we present new observations of the central region of the M~82 starburst obtained with the {\em James Webb Space Telescope} (JWST) near-infrared camera (NIRCam) instrument (Figure \ref{fig:location}), at a resolution ($\theta\sim0.05\arcsec-0.1\arcsec\sim1-2$~pc) and sensitivity unavailable before. These observations are part of Cycle 1 GO program 1701, which also images NGC~253.
Pointing JWST to a very bright source, like a nearby starburst nucleus, carries the risk of saturation. The observations presented here were designed to avoid saturating in the nuclear region and complement larger, deeper mosaics that at the time of writing were not yet observed.
As a new very high resolution view of the archetypal starburst in the near-IR they present a unique opportunity for science.
We discuss the observations and their processing in \S\ref{sec:obs}, we present the results on PAH emission at the base of the outflow in \S\ref{sec:results}, and the conclusions in \S\ref{sec:conclusions}. Companion papers analyze the detailed correlations between outflow tracers (Fisher et al., in prep.), and the identification and properties of massive star clusters (Levy et al., in prep.) based on these same data.

\section{Observations and Data Processing}
\label{sec:obs}

\subsection{Observations}
\begin{figure*}[t]
    \centering
    \includegraphics[width=\textwidth]{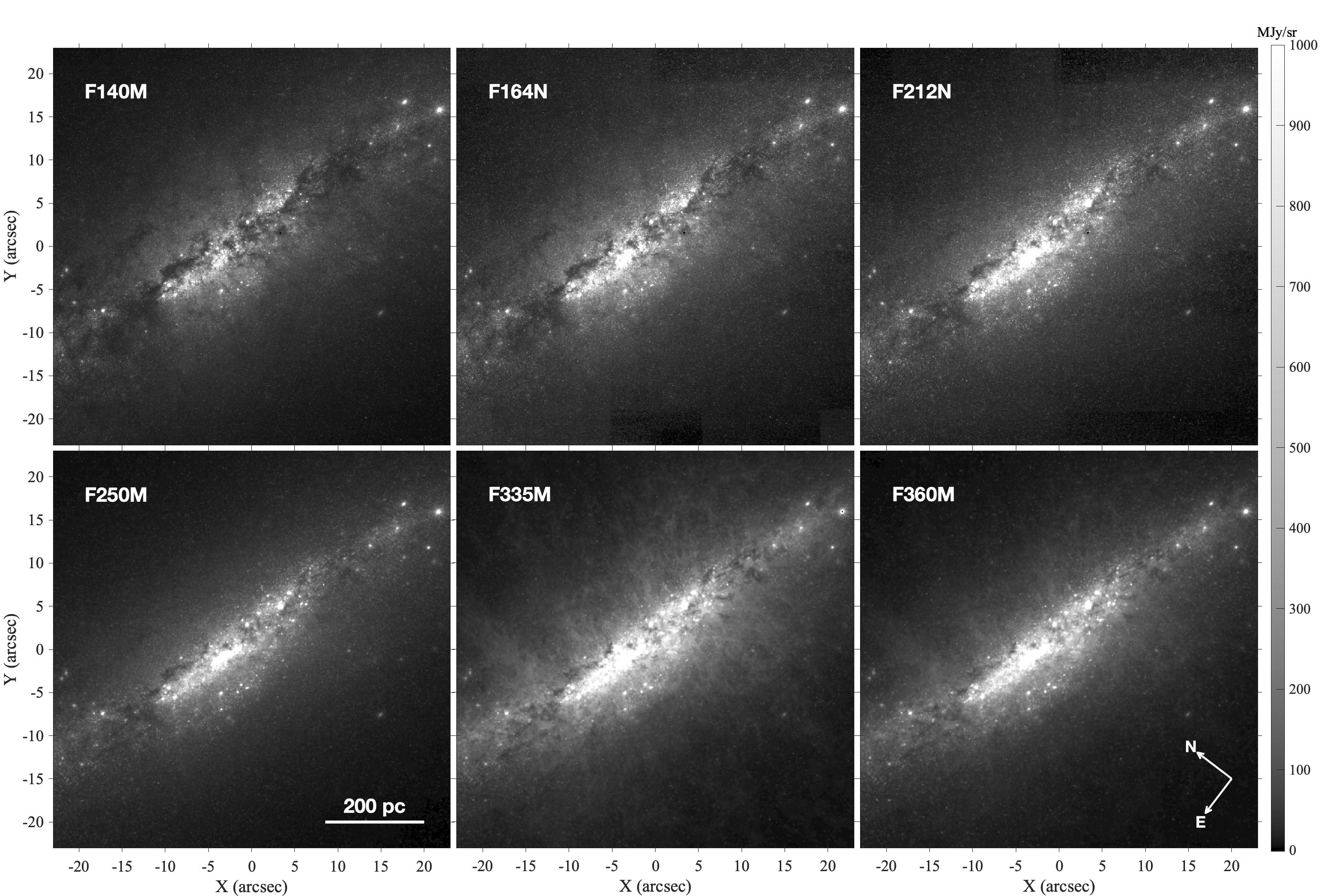}
    \caption{NIRCam mosaics of the center of M~82, with focal plane orientation. The F140M, F250M, and F360M are primarily continuum bands. The F164N, F212N, and F335M are a combination of continuum and emission: F164N covers the 1.64 $\mu$m [FeII] fine structure transition, F212N covers H$_{2}$ vibrational emission at 2.12~$\mu$m, and
    the diffuse emission from the 3.3 $\mu$m aromatic feature is evident in F335M. Contamination by PAH-associated emission can also be seen in F360M. Position offsets are referred to $09^{\rm h}55^{\rm m}51\fs60$, $+69^\circ40\arcmin45\farcs6 $ (J2000). \label{fig:bands}}
\end{figure*}

Observations were acquired by JWST as part of the Cycle 1 GO project 1701 (P.I. Alberto Bolatto) using the NIRCam instrument in a 1.5 hours-long visit starting on October 17, 2022, 23:20 UT. 
To minimize the chance of saturation, the observations were set up using the SUB640 sub-array with the RAPID readout pattern. The mosaics\footnote{Automatic pipeline processed MAST mosaics can be found under \url{http://dx.doi.org/10.17909/cwtn-nh63}}, shown in Figure \ref{fig:bands}, have a linear extent of 50\arcsec\ ($\sim870$~pc) and a resolution of $0.05\arcsec-0.1\arcsec$ ($\sim1-2$~pc).  

\begin{figure*}[t]
\centering
\includegraphics[width=\textwidth]{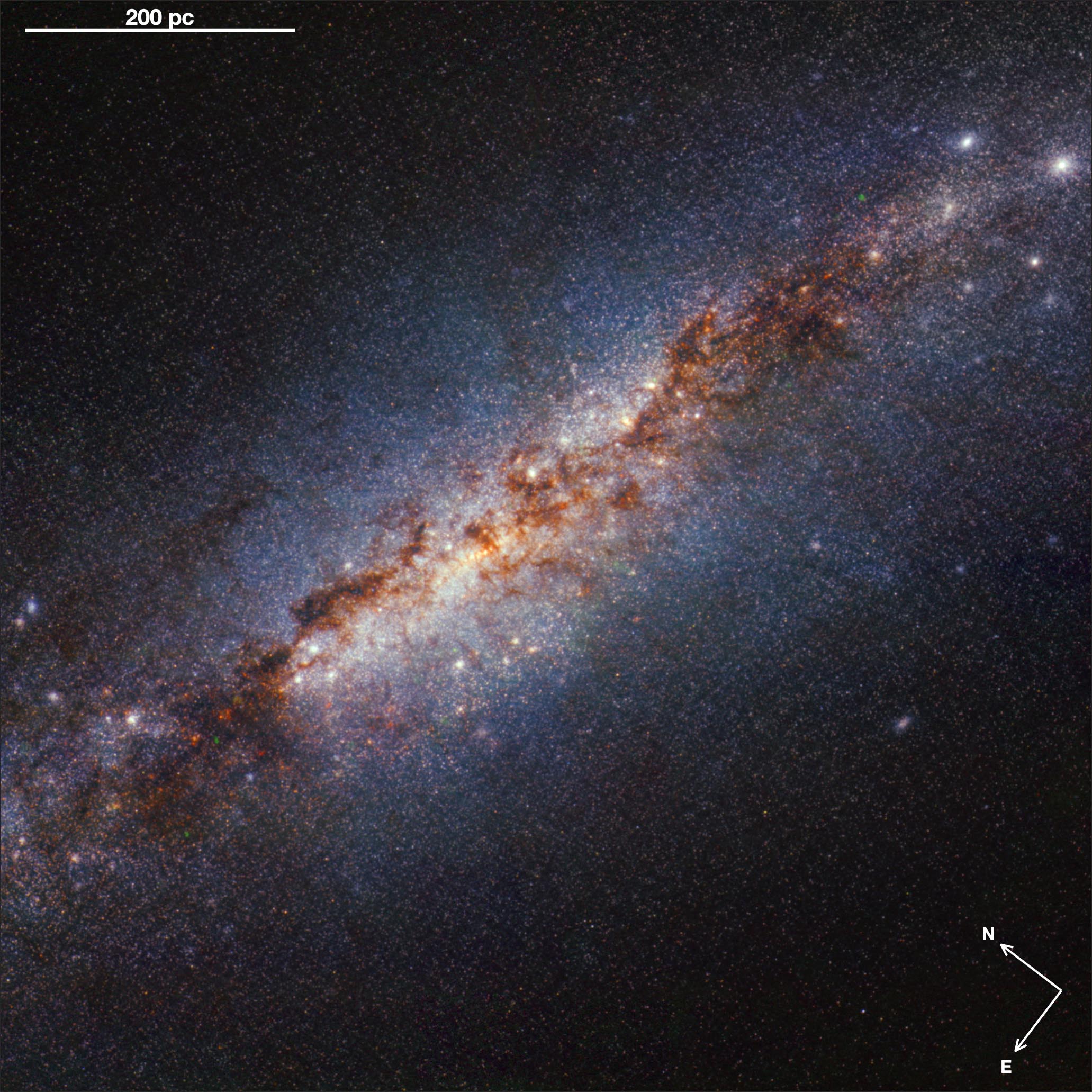}
\caption{Three color image with F212N, F164N, and F140M for red, green, and blue respectively. Some dust entrained in the galactic wind is seen as faint, elongated dark streaks against the bright background. Compact green sources are supernova remnants (see below). Red sources are either highly extinguished or emitting in the vibrationally excited $v=1\rightarrow0$ \htwo, a tracer of shocks and strong UV fields (image credit A. Pagan).  \label{fig:3color1}}
\end{figure*}

\begin{figure*}[t]
\centering
\includegraphics[width=\textwidth]{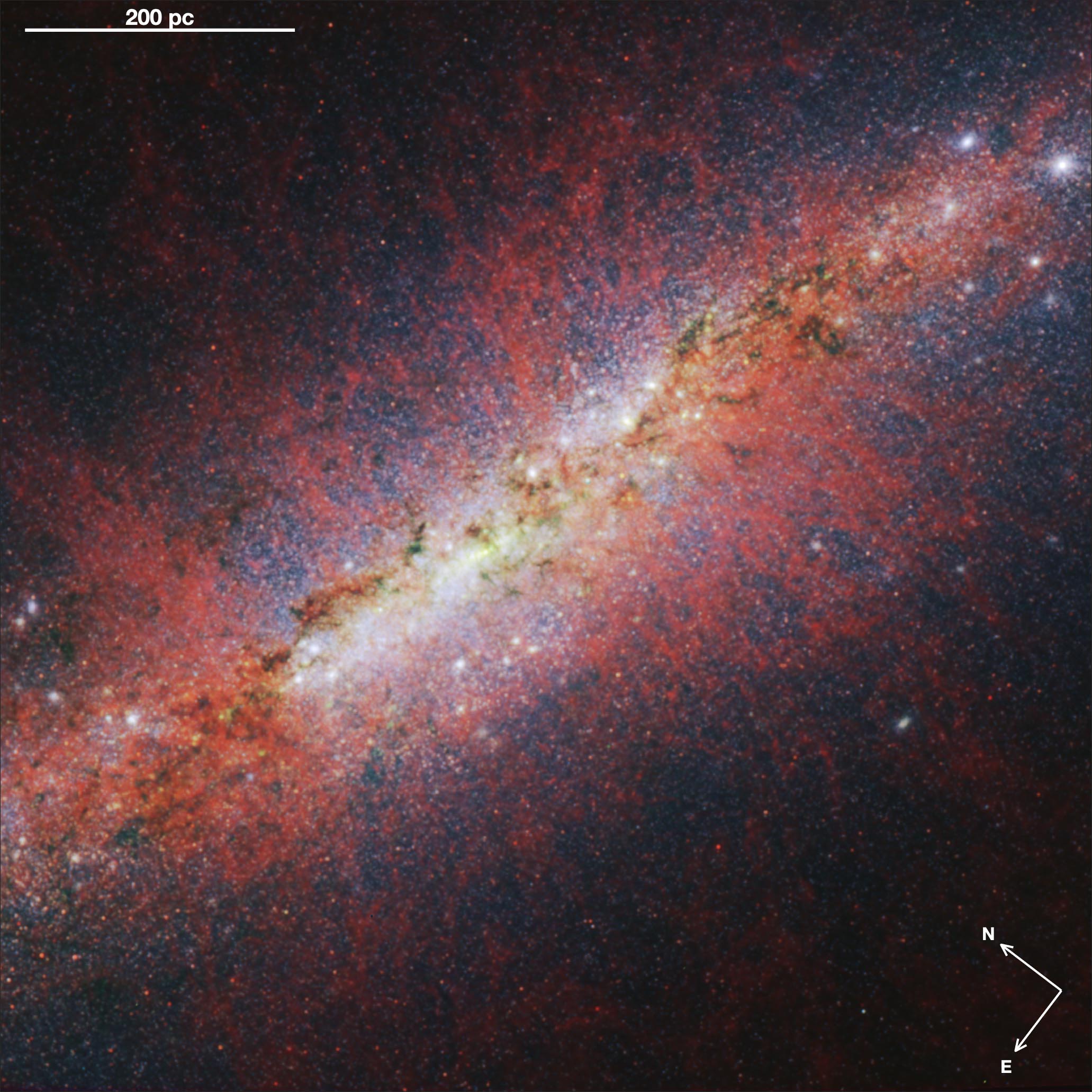}
\caption{Three color image with  F335M, F250M, and F164N  for red, green, and blue respectively. Dust throughout the base of the galactic wind lights up in the 3.35 $\mu$m filter (red) due to PAH emission. The bright F335M emission  appears to outline a truncated bi-cone corresponding to the edge of the outflow base, particularly on the left side ``wall''. In this scenario, the elongated streaks correspond to the front and back surfaces of the bi-cone, seen in projection against the middle regions north and south of the starburst (image credit A. Pagan). \label{fig:3color2}}
\end{figure*}

Data were obtained in six NIRCam filters: F140M, F164N, and F212N on the blue side (SW detector), and F250M, F335M, and F360M on the red side (LW detector). Color combinations of most of these filters are presented in Figures \ref{fig:3color1} and \ref{fig:3color2}.
The F164N and F212N narrow band filters contain a bright fine-structure transition from [\ion{Fe}{2}] and the $v=1-0$ vibrational transition of \htwo\ respectively, while the F335M filter is dominated by a bright 3.3 $\mu$m PAH feature. The mosaic consisted of 4 primary dithers in INTRAMODULEBOX mode and 4 small dithers in SMALL-GRID-DITHER mode, with a depth of 6 groups per integration and a total exposure of just under 470 seconds per filter.

\subsection{Pipeline Processing}
We processed the NIRCam uncalibrated data products with the JWST pipeline version 1.9.6 \citep{Bushouse2023} and CRDS context jwst\_1077.pmap. For stage 1 of the pipeline, we used default parameters except for the following: we used frame0 to recover emission from sources that are saturated in the first integration group in the ramp fitting step (\texttt{suppress\_one\_group = False}). In the jump step, we set \texttt{expand\_large\_events = True} in order to flag the ``snowball'' features that occur due to large cosmic ray events \citep[e.g.,][]{Rieke2023}. We employed all the default parameters for stage 2 processing. In between stage 2 and 3, we applied an additional correction to the \texttt{*\_cal.fits} files for the two narrow-band filters, F212N and F164N, which have the least signal and a comparatively larger contribution from $1/f$ noise from the NIRCam detectors that appears as bright striping features. We adapted the algorithm presented in \citet{Willott2022} for use with the SUB640 sub-array and remove the $1/f$ noise by masking bright sources and subtracting the median of each column.

Aligning the small NIRCam mosaics for M~82 was challenging due to the lack of stars catalogued by the Gaia observatory within the field of view and the small number of bright sources outside of the plane of the galaxy. We used the JWST/HST Alignment Tool \citep[JHAT;][]{Rest2023} to align the individual exposures. First, we aligned the F250M exposures to the stars available in Gaia-DR3 using JHAT. This filter has the least amount of distortion issues because it uses the larger field of view LW detector and contains little to no contamination from PAH emission seen in the F335M and F360M filters. Then, we created a catalog of the stars from the F250M image by selecting pixels with brightness between $1$ and $250$\,MJy\,sr$^{-1}$ (the pixel size is $0.042\arcsec$). This range ensures that the catalog includes the fainter stars outside of the plane of the galaxy, which allows better alignment of the smaller SW detector. We then aligned the rest of the mosaics using this catalog with JHAT. We applied the stage 3 step in the pipeline to the aligned files with the {\tt tweakreg} step turned off, which yielded the final mosaics used in this work. \rev{The $3\sigma$ scatter of the relative positional alignment for all other filters relative to F250M reported by JHAT is 0.03\arcsec\ or better, and the relative alignment between any two filters is typically better than 0.015\arcsec. The absolute astrometry is based on 5 Gaia stars which were found to be in common with sources in F250M, so it is not possible to judge its quality based on the scatter reported by JHAT. In \S\ref{sec:QA}, however, we report the cross-identification of SNRs in the JWST images against VLBI positions, where we find correspondences within 0.2\arcsec. Inspection of the images suggests our absolute astrometry is of order 0.1\arcsec\ or better. Combination of these images with our upcoming large scale, deeper mosaic will enable more extensive cross-matching against Gaia and a better estimate of the absolute astrometric accuracy of the images.} 

\subsection{Point Spread Function Matching}\label{sec:psf_matching}

Matching the Point Spread Function (PSF) of the different filters is necessary for a clean continuum subtraction. We use the {\texttt WebbPSF}\footnote{https://webbpsf.readthedocs.io} tool version 1.2.1 to compute the PSFs of the F250M, F335M, and F360M images. We oversampled the PSFs to obtain a desired pixel size of $0.008\arcsec$. To match the F250M and F335M PSFs to F360M, we generated convolution kernels using {\texttt pypher} \citep{Boucaud2016}. We used the default regularization parameter of $10^{-4}$, which stabilizes the kernel solution by penalizing noisy high frequencies. We experimented with changing this parameter, and it did not affect our results in any significant manner. 

To quantitatively check the performance of these kernels, we adopted the \citet{Aniano2011} ``figure of merit'' ($W_{-}$), which focuses on negative values (see their Equation 21). In essence, the lower the $W_{-}$ value, the less aggressive the kernel. \citet{Aniano2011} find that kernels with $W_{-} \leq 1$ are generally safe to use, though kernels with $W_{-} \approx 1$, while reasonable, are mildly aggressive. The authors caution against using a kernel with $W_{-} > 1.2$, which would be aggressive enough to result in artifacts. For the F250M and F335M matching kernels we found $W_{-} = 0.45$ and $0.98$, respectively. These values indicate that both kernels are reasonably well-behaved and safe to use, with the F335M kernel being only moderately aggressive.

After regridding the images to a common pixel size of $0.008\arcsec$, we performed a fast Fourier transform (FFT) convolution of the F250M and F335M images with the corresponding kernels in order to match the PSFs of the 2.50, 3.35 $\mu$m images to the PSF of the 3.60 $\mu$m image, which has a FWHM resolution of $0.118\arcsec$ (2.06~pc). The clean continuum-subtracted image produced (see next section) validates this PSF matching procedure.

\subsection{Continuum Subtraction}
\label{sec:continuum_subtraction}
To produce an image of the PAH emission band at 3.3 $\mu$m, caused by the C$-$H stretching mode in the aromatic (i.e., benzene-like) rings of PAH molecules, we need to remove the underlying continuum from the F335M image. The 3.3 $\mu$m feature is the brightest of a complex of PAH-related spectral features that extends redwards, including a 3.4 $\mu$m feature thought to be due to C$-$H vibrational modes in aliphatic (chain-like) hydrocarbons \citep{Joblin1996, Lai2020}, and a broad plateau at 3.47 $\mu$m which is observed to be well-correlated with the 3.3 $\mu$m intensity and thought to be also due to aromatic material \citep{Hammonds2015}. The intensities of the 3.3 $\mu$m PAH, 3.4 $\mu$m aliphatic, and the 3.47 $\mu$m plateau are observed to be tightly correlated \citep{Lai2020}. The longer-wavelength features are also in the passband of the F360M filter, which therefore cannot be directly used to estimate the continuum \citep{Sandstrom2023}. In addition, the 3.3 $\mu$m feature sits on the shoulder of a broad water-ice band at 3.05 $\mu$m which sometimes can be seen in absorption in highly extinguished systems, such as luminous and ultra-luminous IR galaxies \citep[e.g.,][]{Imanishi2010,Lai2020}. In contrast to F360M, we expect the F250M filter to be a relatively clean measure of the continuum according to existing spectroscopy of M~82 \citep{Sturm2000,ForsterSchreiber2001}.

Assuming that the contamination of the F360M filter by PAH-related emission is tightly correlated with the PAH-dominated emission detected in the F335M filter \citep[see e.g.,][]{Sandstrom2023}, we can set up an iterative method to separate the PAH and continuum contributions to the signal measured in F335M. In essence we apply the following algorithm:
\begin{enumerate}
\item Compute the continuum signal at 3.60 $\mu$m ($C_{360}$) by removing from the F360M measurement a fraction $q$ of the 3.3 $\mu$m PAH intensity where that intensity is positive.
\item Compute the continuum signal at 3.35 $\mu$m ($C_{335}$) by using a weighted combination of the F250M filter and the 3.60 $\mu$m continuum.
\item Compute the 3.3 $\mu$m PAH intensity by removing the 3.35 $\mu$m continuum from the F335M filter.
\item Repeat until convergence.
\end{enumerate}
\noindent This corresponds to the equations,
\begin{eqnarray}
C_{360} &=& F360M-q \times PAH_{335} \\
C_{335} &=& F250M+(C_{360}-F250M)\times R \\
PAH_{335} &=& F335M-C_{335},
\end{eqnarray}
\noindent where $q$ and $R$ are constants, $PAH_{335}$ is the PAH component in F335M, $C_{335}$ and $C_{360}$ are the continuum component for the F335M and F360M filters, and $q=0$ if $PAH\leq0$. The iteration starts from $PAH_{335}=0$ and produces values for $PAH_{335}$, $C_{335}$, and $C_{360}$. 

Here we will adopt single values for $q$ (the ratio between the contamination in F360M due to aromatic or aliphatic components, and the PAH emission in F335M) and $R$ (yielding the continuum in F335M as a linear combination of the continuum in F250M and F360M) appropriate for our region of interest, the extra-planar emission. Note, however, that in principle these are local quantities with values that change depending on PAH properties and the color of continuum sources.  The typical value of $q$ as determined using the AKARI spectroscopy of entire (unresolved) galaxies in \citet{Lai2020}, is $q=0.36$ (T. Lai, priv. comm.) when including all the features discussed above that contribute to F360M as well as extended wings of bright PAH lines such as the 6.2 $\mu$m feature (just the 3.3 $\mu$m PAH wing by itself would yield $q=0.09$). Our data are well resolved spatially, and the extra-planar region of M~82 contributes little to the integrated light of the galaxy. Its conditions will be substantially different from those in the central regions that dominate the integrated light, so the AKARI result does not necessarily represent it well. The  wispy structure visible in F360M (which is completely absent in the continuum-dominated F250M), however, implies that the PAH contribution is large. This is similar to the findings of \citet{Sandstrom2023} in PAH-dominated regions. 
The factor $R$, on the other hand, captures the fractional contribution of $C_{360}$ to the $C_{335}$ continuum when predicting it via a linear combination of $C_{250}$ and $C_{360}$: for $R=0$, $C_{335}=C_{250}$, while for $R=1$, $C_{335}=C_{360}$. If the slope of the continuum SED is constant between $2.50$ and $3.60$ $\mu$m, we expect $R$ to be close to the ratio of central wavelength differences, $(3.35-2.50)/(3.60-2.50)\approx0.77$, modulo color correction factors.   

\begin{figure*}[t]
    \centering
    \includegraphics[width=\textwidth]{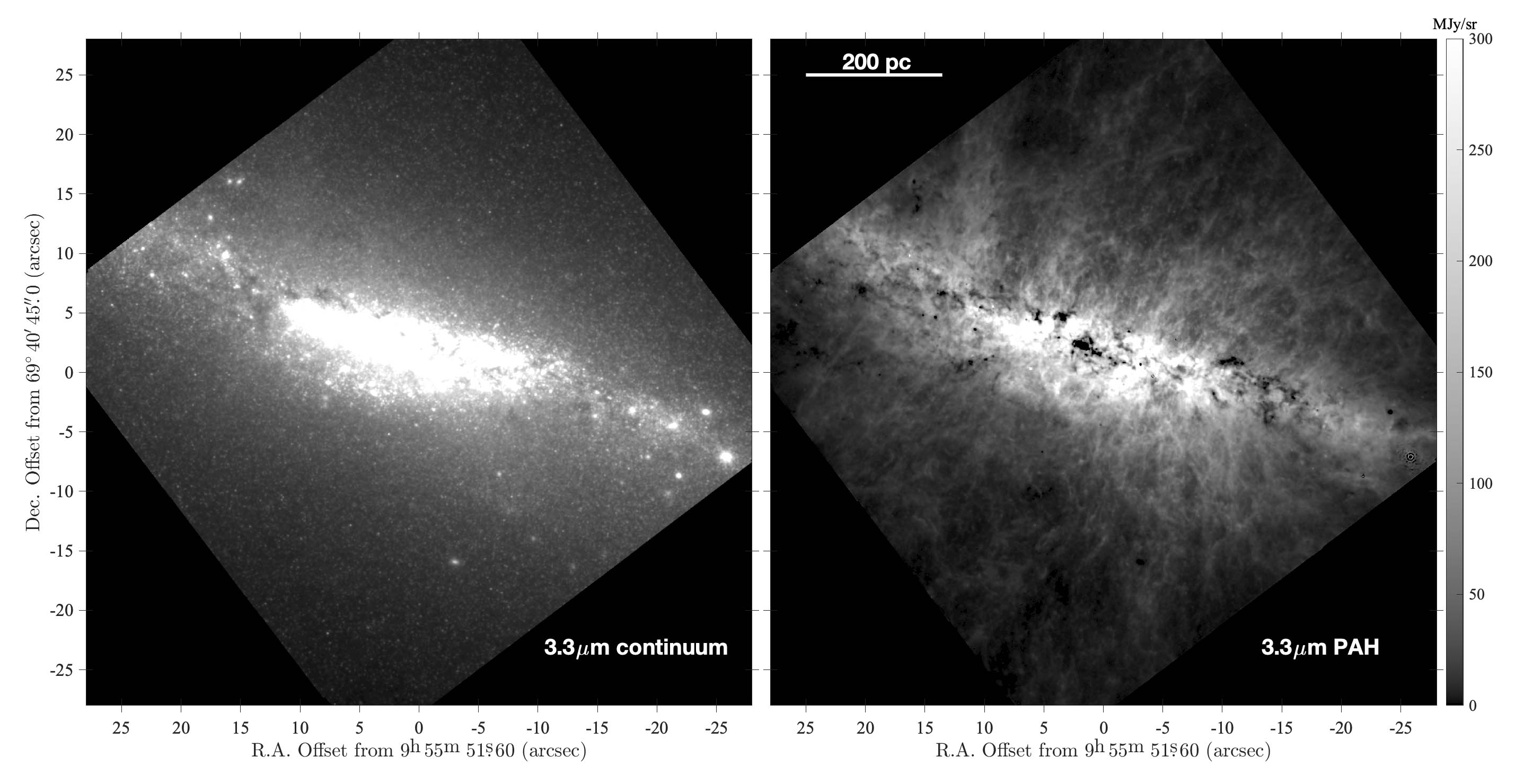}
    \caption{Decomposition of the F335M filter into continuum (left panel) and PAH emission (right panel) following the procedure described in \S\ref{sec:continuum_subtraction}. The extra-planar tendrils of PAH emission are highly structured, and likely represent a view of the cold material entrained in the galactic wind. \rev{Position offsets are referred to $09^{\rm h}55^{\rm m}51\fs60$, $+69^\circ40\arcmin45\farcs0 $ (J2000), the often used center of M\,82 based on 2 $\mu$m ground observations is located at offsets +3.9,+0.8 \citep{Lester1990}}.   \label{fig:pah}}
\end{figure*}

Fortunately, it is possible to arrive at optimal values for $q$ and $R$ by testing a range of values and inspecting the resulting images. The best $q$ for our purposes should result in a $C_{360}$ continuum without the wispy extra-planar structures visible in F360M (Figure \ref{fig:bands}), which are caused by PAH-related emission in this filter, while not introducing negative wispy artifacts caused by oversubstraction from using too large a value for $q$. Similarly, the best value of $R$ minimizes the small scale star-subtraction residuals in the continuum for the extra-planar regions. Using these considerations, we obtain $q=0.45\pm0.05$ and $R=0.55\pm0.05$ (Figure \ref{fig:pah}). The value of $q$ we find is comparable to the $q=0.36$ for entire galaxies from AKARI spectroscopy discussed above, and somewhat smaller than $q\sim0.66$ obtained for very PAH-dominated lines of sight in highly resolved galaxy observations \citep{Sandstrom2023}.

We compare our image with that obtained from the continuum subtraction method used by \citet{Sandstrom2023}, and find both methods produce results largely consistent with each other. We adopt their slope of $B_{\rm PAH} = 1.6$ and offset of $A_{\rm PAH} = -0.2$ for colors in PAH-dominated regions (see their equation 4). While \citet{Sandstrom2023} calculates this slope using the F300M, F335M, and F360M filters, we must substitute the F250M filter for F300M due to availability. This substitution should not drastically change the results, as the \citet{Sandstrom2023} slope appears reasonable in our F335M/F250M${-}$F360M/F250M plane. We find a $\sim10\%$ larger intensity in the $PAH_{335}$ than in the previous method toward the central regions of the map, increasing to $\sim30\%$ in the outer regions. Most importantly, the overall morphology of the extended PAH features remains unchanged when adopting the \citet{Sandstrom2023} approach.

\subsection{Ancillary Data}
\label{sec:ancillary}

A 6 GHz image from Marvil et al. (in prep.) is derived from data taken with the Karl G. Jansky Very Large Array (VLA) between Feb 2011 and March 2012 in antenna configurations A  (5 hours), B (10 hours) and C (3 hours) under project codes TDEM0010, 10C-199 and 12A-457, respectively.  The data were processed in CASA \citep{CASATeam2022},  using the flux-density scale of \citet{Perley2013} for calibration and the multi-scale, multi-frequency synthesis algorithm \citep{Rau2011} for imaging. The 6 GHz continuum image combines 2048 x 2 MHz channels spanning the entire 4-8 GHz frequency response of the C-band receiver. The synthesized beam has a FWHM of 0.36\arcsec, and the RMS noise level of the image is $3.5$\,$\mu$Jy/beam. Although the image includes short spacing information from the compact Jansky VLA configurations, the galaxy emission sits on a mild negative bowl of order $\sim 5$\,$\mu$Jy/beam that suggests some extended emission is missing, but this does not affect our calculations.

The Paschen $\alpha$ observations are from the Mikulski Archive for Space Telescopes (MAST) and were obtained by the Hubble Space Telescope (HST) with the Near Infrared Camera and Multi-Object Spectrometer (NICMOS) with two filters: F187N (PID 7919, P.I. W. Sparks), and F166N (PID 7218, P.I. M. Rieke)\footnote{\url{http://dx.doi.org/10.17909/jdx7-qg88}}. The continuum subtraction for the Paschen~$\alpha$ image uses F166N and is described by Fisher et al.\ (in prep.). To briefly summarize, the subtraction followed the procedure outlined in \cite{Boker1999}, in which the flux in F166N is scaled to the flux of F187N, with a scaling determined in regions avoiding the bright midplane of the galaxy. For a given F166N flux, there is a well-defined minimum of the F187N flux. Pixels with values near the minimum F187N/F166N trace the ratio of stellar continuum in the two filters, while pixels with larger F187N/F166N are dominated by Paschen~$\alpha$ emission. To determine the continuum, we fit a linear correlation between F187N and F166N for those pixels in the lowest 0.5\% of the ratio F187N/F166N, which is then subtracted from all pixels in the F187N image. 

The CO \jone\ image from \citet{Krieger2021} is part of a large mosaic obtained using the Institut de Radioastronomie Millim\'etrique (IRAM) Northern Extended Millimeter Array interferometer (NOEMA). It includes short spacings from the IRAM 30m telescope, so in principle it recovers all the source flux. It has a synthesized beam of $2.08\arcsec\times1.65\arcsec$ (i.e., a circular beam equivalent $\theta\approx1.85\arcsec$), and an RMS sensitivity of 138 mK (5.15\,mJy/beam) in a 5\,\kmpers\ channel.

\begin{figure*}
    \centering
    \includegraphics[width=\textwidth]{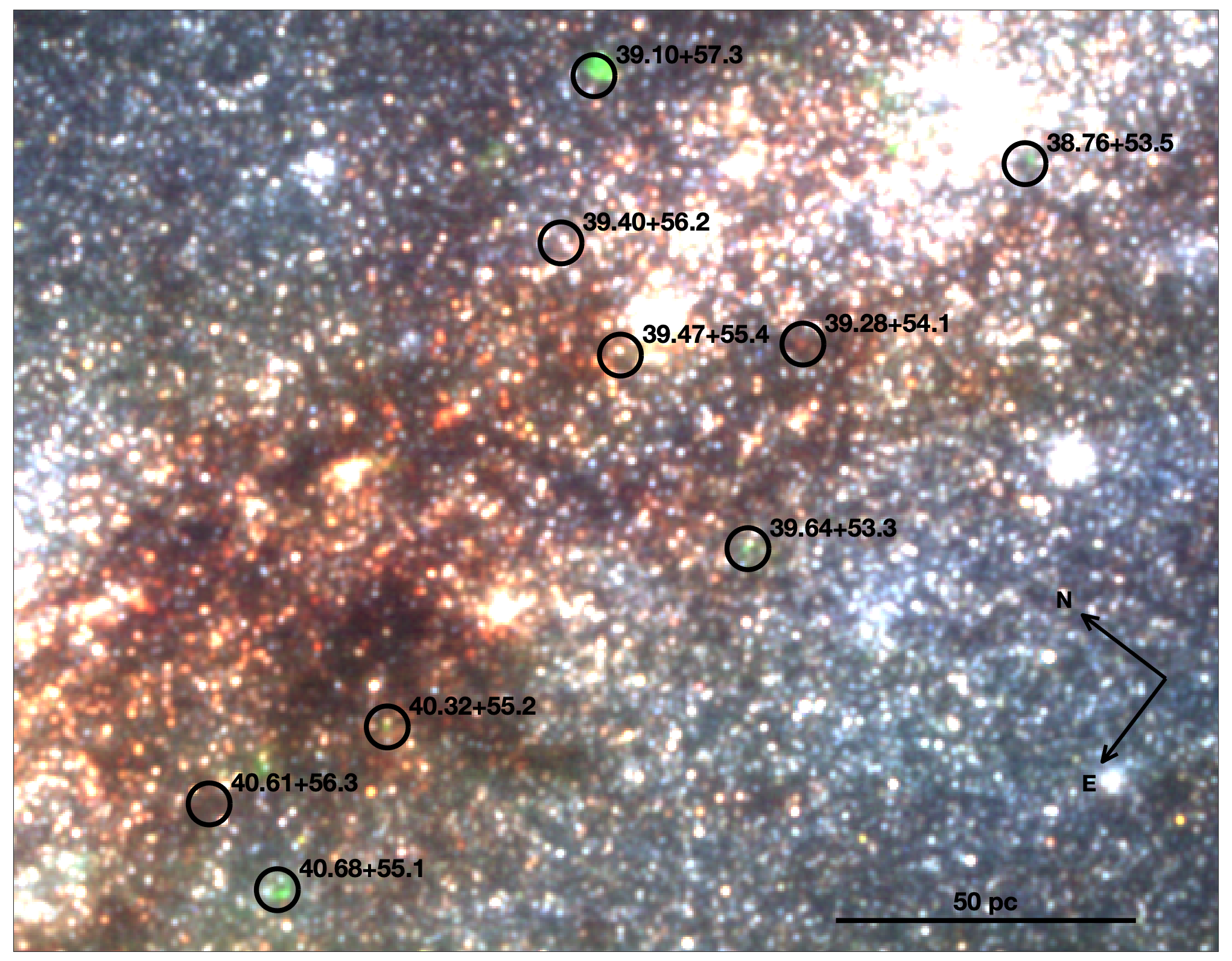}
    \caption{A $11.5\arcsec\times9\arcsec$ ($200\times160$~pc) area near the west corner of the mosaic, with radio SNR positions overlaid (the orientation is the same as Figure \ref{fig:bands}) The RGB NIRCam mosaic uses the F212N, F164N, and F140M filters for red, green, and blue respectively. The circles (radius 0.2\arcsec, or 3.5\,pc) and labels correspond to the radio SNRs  \citep[][Table 5]{Fenech2010}. Green regions show an excess of emission at 1.64 $\mu$m likely associated with the [\ion{Fe}{2}] line due to dust destruction in supernovae shocks. This excess is visible on the radio SNRs 39.10+57.3, 40.68+55.1, 39.64+53.3, 38.76+53.5, and 40.32+55.2, although some regions not associated with known radio SNRs also show a possible F164N excess, perhaps due to local or large-scale shocks. Very red sources are either heavily extinguished at 2 $\mu$m or associated with vibrational emission from \htwo, usually caused by shocks.\label{fig:SNRs}}
\end{figure*}

\begin{figure*}[t]
    \centering
    \includegraphics[height=0.9\textheight]{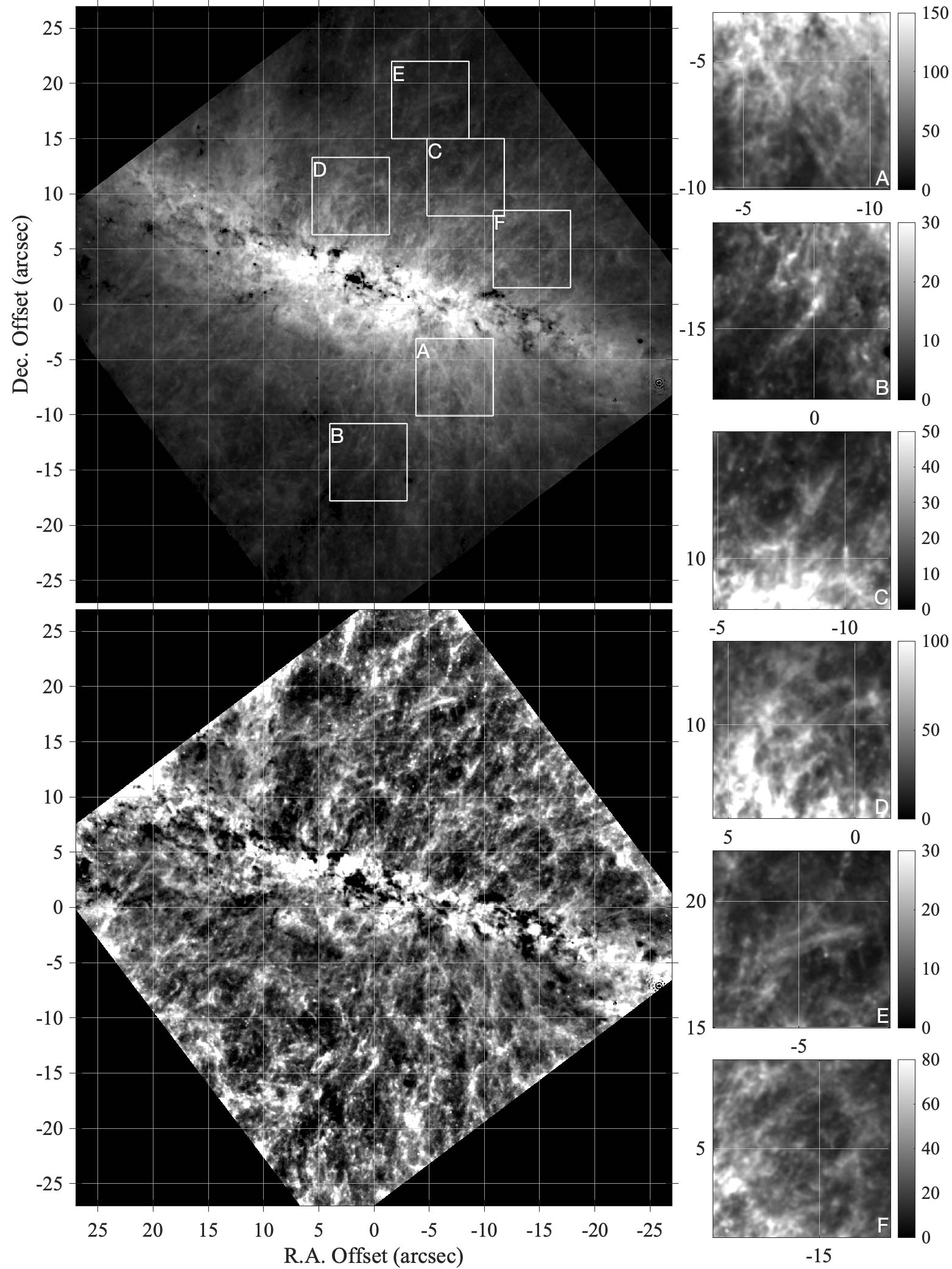}
    \caption{Small scale structure in the PAH emission. The top panel shows the same image as Figure \ref{fig:pah} with a grid superimposed. The bottom panel shows the result of removing a version of the image smoothed on 5\arcsec\ (87~pc) scales, a technique known as unsharp masking, to bring up the structure below those scales. This highlights the network of fine filaments and the bright walls of bubble-like structures present in the PAH image. The panels to the right show $7\arcsec\times7\arcsec$ zooms into the top panel, highlighting the rich substructure in the outflow. Their gray scale is linear and adjusted separately for each panel, as indicated by the accompanying colorbar (in MJy/sr). \rev{Reference position is as in Figure \ref{fig:pah}.}      \label{fig:unsharp}}
\end{figure*}

\section{Results and Discussion}
\label{sec:results}

\subsection{Multi-Wavelength Images of the Starburst}
\label{sec:QA}
The wavelength progression in Figure \ref{fig:bands} shows clearly the dramatic reduction in extinction at longer wavelengths. One of the most striking features of these mosaics are the highly structured extra-planar tendrils of emission seen in the F335M and F360M filters. These are associated with PAH emission features around 3.3 and 3.4 $\mu$m, as discussed in \S\ref{sec:continuum_subtraction}. 

A three-color combination of near-IR data in Figure \ref{fig:3color1}  shows images from our three filters on the blue side of NIRCam (1.40, 1.64, and 2.12 $\mu$m), maximizing the image resolution. This high resolution allows us to identify resolved stellar sources corresponding to clusters with stellar masses $\gtrsim10^4$~\msun\ and larger, which are described in a companion paper (R. Levy et al., in prep.). The highly obscured starburst region is partially visible at these wavelengths, along with dark extra-planar filaments emanating from the central region due to dust entrained in the galaxy wind. 
Figure \ref{fig:3color2} uses the combination of the 1.64, 2.50, and 3.35 $\mu$m filters to show the PAH emission associated with the entrained dust. We discuss the structure of the emission, which we separate from the continuum as described in \S\ref{sec:continuum_subtraction} (Figure \ref{fig:pah}), in \S\ref{sec:structure}. \rev{Note that the bright extra-planar structures seen in PAH emission in Figure \ref{fig:3color2} have corresponding extinction (dark) features due to dust lanes in Figure \ref{fig:3color1}, seen against the blue background of stars. In particular, the left side ``wall'' of PAH emission and some of the prominent PAH plumes on the north side show clear corresponding extinction features. }

The bright green compact sources in Figure \ref{fig:3color1} are regions with enhanced 1.64 $\mu$m emission. They mostly correspond to supernova remnants (SNRs), visible as an excess in our F164N image due to the contribution from the [FeII] transition at 1.644 $\mu$m, due to shock-induced destruction of dust grains and the subsequent release of iron atoms into a gaseous phase. The first extragalactic source detected in the 1.644 $\mu$m [FeII] transition was indeed M~82 \citep{Rieke1980}. SNRs also emit brightly in the radio due to synchrotron emission powered by relativistic electrons accelerated by shocks \citep{Dubner2015}. We can thus compare our 1.64 $\mu$m excess positions with radio catalogued SNRs, in particular the catalog from 1.7 GHz observations with the Multi-Element Radio Linked Interferometer Network (MERLIN) and Very Large Baseline Interferometry (VLBI) observed by \citet[][Table 5]{Fenech2010}.

Figure \ref{fig:SNRs} shows a zoomed area in detail, illustrating the very good agreement in several SNe between the [\ion{Fe}{2}] emission and radio data (black circles). The SNRs visible at 1.64 $\mu$m are within 0.2\arcsec\ of their radio positions, providing an independent check of the absolute astrometric accuracy. Not all radio SNRs have clear associated [FeII] emission. Some of them are located in regions that are very heavily extinguished at 2 $\mu$m, and that may be the cause. Conversely, not all regions with excess in F164N correspond to catalogued radio SNRs. Some of these regions may correspond to unidentified SNRs or extended shocks associated with the galactic wind, a matter of future research.  

\begin{figure*}[t]
    \centering
    \includegraphics[width=\textwidth]{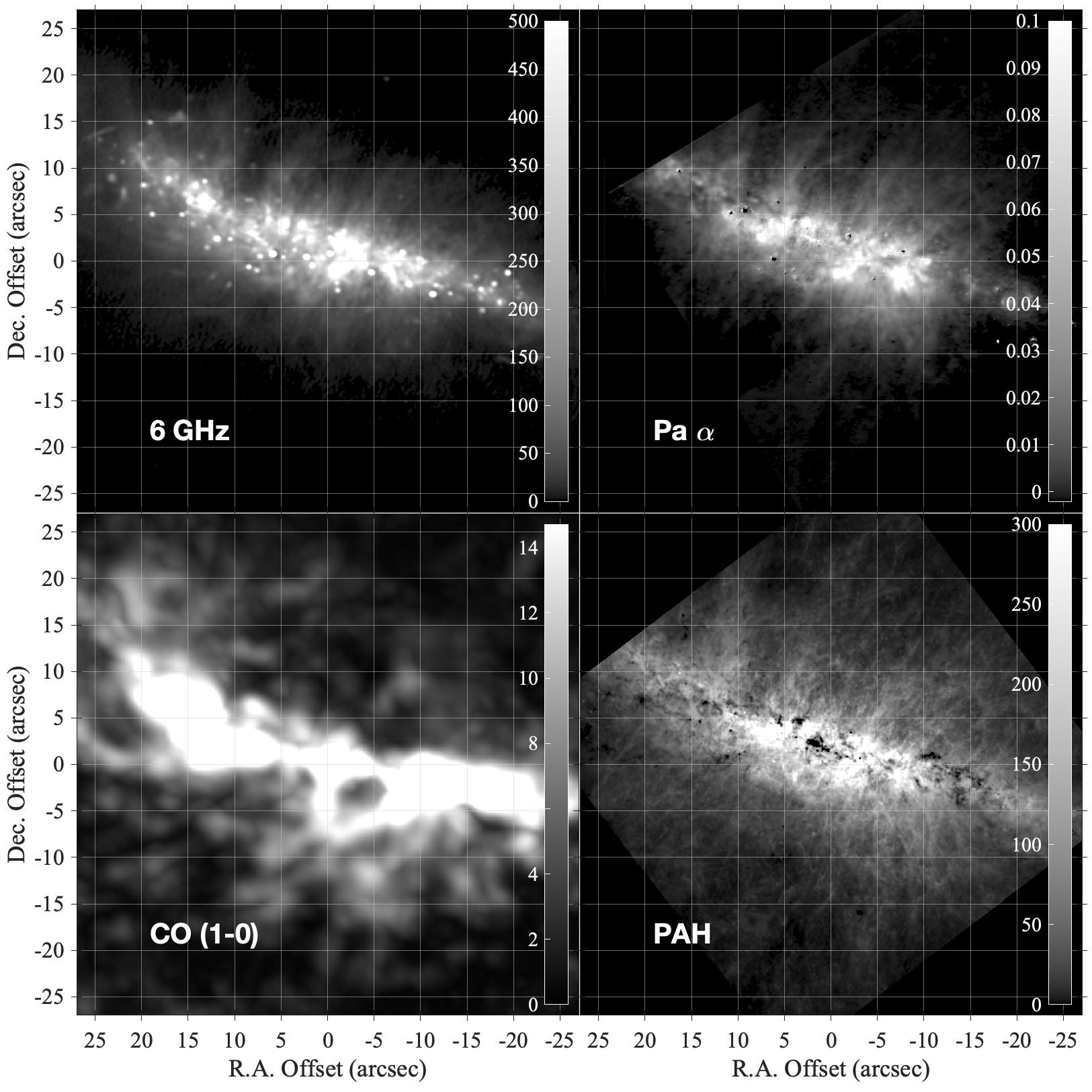}
    \caption{Comparison of different tracers of extra-planar material. A 5\arcsec\ scale corresponds to 87.3\,pc at the distance of M~82. Top left: Jansky VLA 6 GHz continuum at 0.36\arcsec\ ($6.3$\,pc) resolution (square root stretch, units MJy/sr). Top right: HST archival continuum-subtracted Paschen $\alpha$ at $0.25\arcsec$ ($4.4$\,pc) resolution (square root stretch, units erg\,s$^{-1}
    $\,cm$^{-2}$\,sr$^{-1}$). Bottom left: NOEMA CO $1-0$ peak intensity from \citet{Krieger2021} at $1.85\arcsec$ ($32.3$\,pc) resolution (linear stretch, units K). Bottom right: JWST 3.3\,$\mu$m PAH emission at 0.11\arcsec\ ($1.9$\,pc) resolution (square root stretch, units MJy/sr). The ratio of 6 GHz to Paschen $\alpha$ implies the extra-planar filamentary radio continuum emission is thermal (see \S\ref{sec:structure}) The PAH emission shows very good detailed correspondence with the ionized gas traced by Paschen $\alpha$ and the radio continuum. A faint grid is superimposed to facilitate comparison. \rev{Reference position is as in Figure \ref{fig:pah}.}\label{fig:tracers}}
\end{figure*}

\subsection{Extra-planar PAH Emission and the Structure of the Wind}
\label{sec:structure}

The combined data in Figure \ref{fig:3color2} shows that material entrained in the wind emits in the PAH near-IR bands. We produce an image of the 3.3~$\mu$m PAH emission using the procedure described in \S\ref{sec:continuum_subtraction} to separate the continuum and the PAH contributions to the F335M image (Figure \ref{fig:pah}). The separation works very well, with no wispy features or artifacts present in the continuum image. The 3.3 $\mu$m PAH emission has almost no stellar residuals, although there are negative (black) areas in the central regions of the starburst. These areas tend to correspond to highly-reddened, highly extinguished dark clouds, and they are mostly mildly negative ($\sim -30 $ MJy/sr) except in the area at the very center of the starburst (near offsets +2,+2) where it becomes deeply negative ($\sim -1000$ MJy/sr). Although the mildly negative regions depend on the details of the continuum subtraction (i.e., the local value of $q$) the central deeply negative region does not: the F335M filter there is considerably below the baseline determined by the continuum at 2.50 and 3.60 $\mu$m. As pointed out above, the F335M filter is on the shoulder of a broad water-ice feature centered at 3.05 $\mu$m, so the simplest way to make it underluminous is to have a deep ice absorption feature. But it is also possible that the relative mix of aliphatic (which contribute mostly to the F360M filter) and aromatic components changes drastically in some of these regions. A large growth in the aliphatic component  \citep[expected to be higher in dense regions,][]{Rau2019} at these locations may raise preferentially the emission in the F360M filter. We hypothesize that very negative regions are mostly due to water ice absorption along heavily extinguished lines of sight, but definitive confirmation will require spectroscopy. 

The PAH image shows prominent plumes or pillars of PAH emission extending outward from the central starburst region, together with a network of complex filamentary substructure and edge-brightened bubble-like features. An example of one of these plumes (which appear to be bundles of filaments) is the structure North of the mid-plane at R.A. +10, which extends roughly N-S to the edge of the image and forming the left segment of a V shaped pattern with a neighboring plume. Extended emission from dust grains and PAHs associated with the wind has been detected before in M~82, over scales of 6~kpc \citep{Engelbracht2006,Beirao2015}. Our imaging extends to $\sim0.4$~kpc from the central region, representing only the innermost part of the material in the wind but providing a much sharper view than was possible with the {\em Spitzer Space Telescope}. 

We can appreciate better the detailed structure of the PAH emission in Figure \ref{fig:unsharp}, where the bottom panel presents an unsharp-masked image that brings up the contrast on the small scales. To obtain this image we performed unsharp masking in logarithmic space through the following steps: 1) clip the original image $I$ (top panel) below zero, 2) take the $\log(1+I)$, 3) convolve with a 5\arcsec\ (87~pc) FWHM Gaussian kernel to produce $I_S$ (this scale was chosen so that it preserves reasonably well the large scale structure of the ``plumes'' while enhancing scales smaller than 100~pc), 4) compute the final image $I_{UM}=[(1+I)/10^{I_S}]-1$, which is displayed on a scale of $-0.5$ to $1$. The insets in Figure \ref{fig:unsharp} provide a zoom into some of the wind substructure. The image shows a network of intertwined filaments, which appear as coherent structures with lengths $\sim100$~pc in many cases, and thicknesses \rev{of $5-9$~pc} approaching the 1.9~pc resolution of the image. \rev{Qualitatively similar structures are created in simulations by the destruction of dense clumps of material immersed in a wind \citep[e.g.,][]{Schneider2020,Abruzzo2022,Fielding2022}, where cold gas is ablated from clumps and extended into filamentary structures that survive due to stabilization caused by fast cooling.}  

Particularly striking are structures with the morphology of elongated bubble walls, forming a ridge north of the mid-plane and parallel to it, on the west side of the image (e.g., Figure \ref{fig:unsharp} panel F). The larger bubbles on this part of the image extend to $\sim150$~pc from the mid-plane. The north and south sides of the outflow look somewhat different, with the north (the receding side) having more defined large-scale structures than the south. 
The bubble-like structures seen in PAH emission are also apparent in high-resolution H$\alpha$ images, particularly in the less extinguished south (approaching) side of the wind cone (Figure \ref{fig:location}). They may arise from super-bubbles pumped by individual SNe events ejecting material. Another possibility is that they are due to wind instabilities. 
For example, if the wind in the warm gas is driven by pressure of streaming cosmic rays, the characteristic growth time
scale of instabilities would be $\tau_{\rm grow}\sim H/c_{cr}$, where $c_{cr}^2=(2/3)P_{cr}/\rho$ for $P_{cr}$ the cosmic ray pressure, $\rho$ the density of the gas, and $H$ is a pressure scale-height \citep{Quataert2022a}. The production of structures by the instability requires that the growth time is shorter than the flow timescale determined by the wind speed, $v_w$, so that  $\tau_{flow}\sim H/v_w$, 
requiring $c_{cr}>v_w$, which in turn may be attainable for moderate $P_{cr}$ if $\rho$ is not large. 

In Figure~\ref{fig:tracers} we compare the JWST PAH emission with the highest resolution available images of: 1) 6 GHz radio continuum from the VLA (Marvil et al., in prep.), 2) ionized gas Paschen $\alpha$ recombination emission from HST (Fisher et al, in prep.), and 3) molecular gas from NOEMA \citep{Krieger2021}. See  \S\ref{sec:ancillary} for more details on these ancillary data. The similarity of the extra-planar filamentary features present in both the 6 GHz and Paschen $\alpha$ images is particularly striking, and the same broad streams of filaments are present in the PAH 3.3 $\mu$m emission. 
\rev{There is a recombination transition at 3.297\,$\mu$m (Pfund~$\delta$) that could contribute to the line flux in F335M, and directly trace ionized emission potentially confusing the interpretation of the observations. Calculation shows, however, that it can only account for $\sim0.5$~MJy\,sr$^{-1}$ (out of the several tens of MJy\,sr$^{-1}$ present in the PAH filamentary structures) using the observed Paschen~$\alpha$ flux and the relative intensity coefficients for case B recombination at low density by \citet{Hummer1987}.}

For comparison, in the Milky Way center the longest radio filaments (which can be as long as $\sim100$~pc) are caused by non-thermal synchrotron emission at 1.28 GHz, and appear to be related to a wind from SgrA$^*$ \citep{Yusef-Zadeh2023}. In M~82, however, the  ratio of the 6 GHz continuum to recombination line emission we measure for gas off the plane of the galaxy is $T_b({\rm mK})/I_{H\alpha}({\rm Rayleigh})\approx0.1-0.2$, assuming case B recombination to convert Paschen $\alpha$ to H$\alpha$ \citep{Osterbrock2006}. This approximately corresponds to the expected ratio for thermal free-free emission from plasma at about $T_e\approx 5,000$~K \citep[$T_b({\rm mK})/I_{H\alpha}({\rm Rayleigh})\approx0.18$,][Eq. 1]{Gaustad2001} assuming obscuration is unimportant for Paschen $\alpha$ (an extinction correction would drive the ratio even more into the thermal regime for the 6 GHz emission, and a reflection component would act in the opposite direction). 
This suggests that the elongated filamentary structures are not due to a very strong magnetic field, which presumably would cause the radio emission to be dominated by synchrotron radiation even at 6 GHz and yield a non-thermal ratio.  
\rev{Could the emission from these structures be dominated by synchrotron at lower frequencies near 1 GHz, where such measurements are typically made? For a synchrotron spectral index of $\nu^{-0.7}$ the emission would grow by a factor of 3 between 6 GHz and 1.3 GHz. This suggests that if synchrotron were contributing just under 40\% of the radio emission at 6 GHz, it could become just over 60\% of the emission at 1.3 GHz. Therefore, our measurements at 6 GHz place quite a strong constraint on the emission contributed by non-thermal electrons gyrating in magnetic fields associated with the base of the M\,82 outflow.}

\begin{figure}[t]
\centering
\includegraphics[width=\columnwidth]{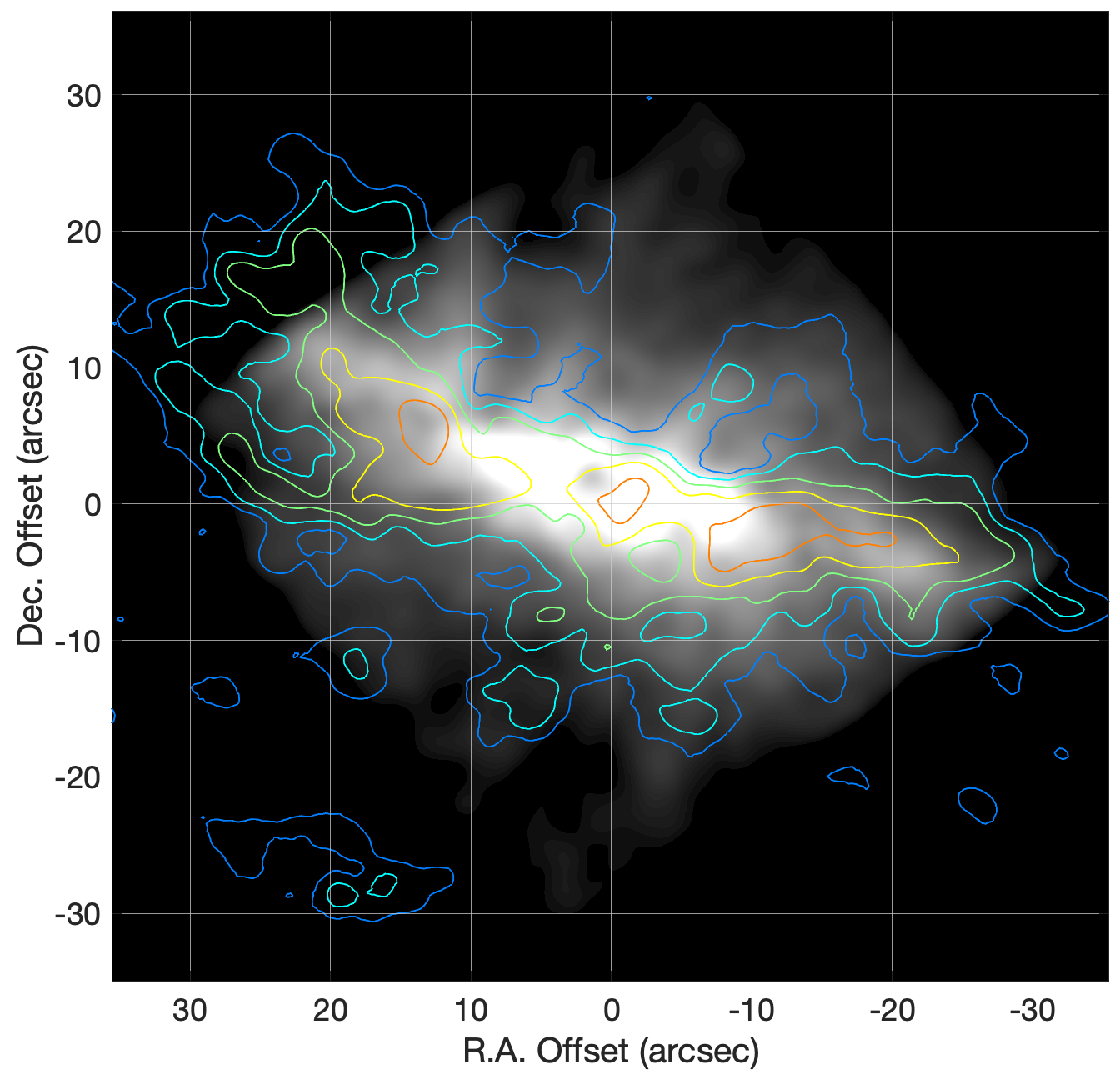}
\caption{PAH and CO emission compared at the same resolution. The JWST 3.3 $\mu$m PAH emission (gray scale) has been convolved to the 1.85\arcsec\ resolution of the NOEMA CO, presented here as the integrated intensity map (color contours). The contour values are 190, 320, 550, 930, and 1580 \Kkmpers. The lowest contour corresponds to a column density $N(\htwo)\sim2\times10^{22}$\,\percmsq. Partial correspondence is observed between the base of some of the PAH plumes and extra-planar CO emission. \rev{Reference position is as in Figure \ref{fig:pah}.} \label{fig:CO_PAH}}
\end{figure}

By contrast, even after accounting for the different resolution, there is just a weak correlation between the extraplanar structures in the PAH emission and those in the molecular gas image. This is also true when comparing with an integrated intensity map, instead of the peak intensity shown in Figure~\ref{fig:tracers}.  It is hard to interferometrically image faint emission in the vicinity of a bright source, especially with a small number of baselines.  We tested the quality of the faint extra-planar CO emission by comparing the CO $1-0$ mosaic with a very high resolution CO $3-2$ mosaic obtained by the Submillimeter Array (SMA, Jim\'enez-Donaire et al., in prep.) and found reasonably good correspondence between the faint emission in both CO maps, suggesting that the imaging quality for the faint extra-planar gas is not the cause of the lesser degree of correlation between PAH and CO emission. 

The correlation between PAH and CO emission is poor, but not completely absent. Since the resolution of the CO data is 17 times lower than that of the PAH image we only expect large scale features to be correlated, and indeed some of them are present both in the CO and the PAH emission. For example, the ``wall'' of CO emission that starts at +10\arcsec\ R.A., and larger offsets on the north-east corner of the image, matches the location of the plume of PAH emission discussed previously. Figure \ref{fig:CO_PAH} shows the matched resolution PAH image with CO integrated intensity contours overlaid to facilitate comparison. The base of some of the PAH plumes, in both the northern and southern outflows, clearly correlates with CO emission. Note that the lowest CO contour still represents a fairly high column density, $N(\htwo)\sim2\times10^{22}$\,\percmsq\ (computed assuming $\xco\sim1\times10^{20}$\,\xcounits), and this may be at least in part responsible for the poor correlation. These are column densities of giant molecular clouds in the Milky Way, so what is seen in CO are substantial clouds. Quantitative comparison of spatial profiles of emission at different heights from the plane is presented in Fisher et al. (in prep.). 

\subsection{Origin of the PAH emission}

Presumably, the PAH emission is tracing the cooler gas entrained by the hot outflow, since small dust grains in hot, X-ray emitting gas should be very quickly destroyed by sputtering caused by collisions with fast-moving electrons \citep{Micelotta2010b,Hu2019}. This is more so for PAHs emitting at 3.3 $\mu$m, a feature that is thought to be dominated by the smallest PAHs \citep{Draine2021}, which are those most quickly ablated by sputtering \rev{and fragmented by photodestruction processes}. 

The timescale to travel from the mid-plane to half the vertical extent of the extra-planar emission in our mosaic (a distance of $\sim200$~pc) at a typical velocity for a neutral or molecular gas wind \citep[$v_{w}\sim200-300$~\kmpers,][]{Leroy2015b,Yoshida2019} is $\tau_{\rm dyn}\sim1$~Myr, and the shortest possible $\tau_{\rm dyn}\sim0.2$~Myr would correspond to gas moving at the velocity of the hot phase ($v_w\sim1000$~\kmpers). By comparison, the expected time for sputtering a 50-atom PAH immersed in T$\approx6\times10^6$~K gas with density $n\approx10^{-2}$~\percmcu\ is $\tau_{\rm sputter}\sim10^{-3}$~Myr, and even 200-atom PAHs have predicted lifetimes $\tau_{\rm sputter}\lesssim10^{-2}$~Myr \citep{Micelotta2010b}. The central region of M~82 may have even hotter ($T\approx3\times10^7$~K) and denser ($n\approx0.1$~\percmcu) conditions \citep{Lopez2020}, which would only accelerate the PAH destruction. Therefore it appears impossible for the PAHs responsible for the observed emission to be mixed with hot, X-ray emitting gas.  

As a consequence we expect the PAHs to be associated with colder phases, either neutral atomic/molecular gas or possibly photoionized gas with $T\sim3\times10^3-10^4$~K. The latter represents a gas phase in which PAHs are not usually observed. PAHs appear to be faint or absent in the ionized gas of \hii\ regions \citep{Chastenet2019,Chastenet2023,Egorov2023}. High resolution studies of the Orion Bar photodissociation region (PDR) show a sharp increase in the brightness of the PAH emission going from the \hii\ region into the molecular cloud and the neutral atomic layer surrounding it, at the location of the ionization front \citep{Peeters2023}. Although PAH emission is seen in the ionized gas region, it is attributed to a background PDR. 
\rev{Observations of the Horsehead nebula suggest a lower limit of $5\times10^3$~yr for the lifetime of PAHs in ionized gas \citep{Compiegne2007},}
but in general their destruction timescale is highly uncertain. The collisional destruction calculations by \citet{Micelotta2010b} indicate that PAHs embedded in high density ($n\sim10^4$\,\percmcu) photoionized gas are destroyed mostly by impacts from He nuclei, and their expected lifetime is $\tau_{\rm sputter}\sim10$~Myr. Since the destruction rate is proportional to the frequency of collisions, \rev{$\tau_{\rm sputter}\propto n^{-1}$}. Therefore the PAH lifetime to collisional destruction will be even longer at a more reasonable density of $n\lesssim10^3$\,\percmcu\ for the base of the outflow \citep[measurements indicate $n\sim200$\,\percmcu\ 500~pc away from the midplane, increasing inward,][]{Xu2023}. These timescales are well in excess of the dynamical time $\tau_{\rm dyn}$ needed to travel over the extent of our image, which makes it plausible that some fraction of the PAH emission originates from photoionized gas, if collisions are the primary destruction mechanism. Photodestruction \citep[e.g.,][]{Jochims1999,Tielens2008}, which preferentially removes the smallest PAHs \citep{Allain1996,LePage2003}, may be however the ultimate mechanism limiting their lifetime as it likely occurs for \ion{H}{2} regions. 
\rev{Neutral or molecular gas advected from high density regions in the walls of the outflow and recently exposed to ionizing radiation could act as a source of PAHs constantly replenishing the ionized gas.} 
Also, it is possible that at least part of the PAHs observed are related to large dust grain shattering in shocks, which can tilt the dust grain size distribution toward smaller sizes \citep{Jones1996,Seok2014}, and provide a PAH source. \rev{Note that PAHs are rapidly destroyed in shocks with velocities $\gtrsim100$~\kmpers\ \citep{Micelotta2010a}, therefore a net PAH generation would require slower velocities. Shocks associated with the interface between hot and warm gas \citep{Fielding2022} would be very inhospitable to PAHs.}

Neutral gas harboring PAHs in the extra-planar region of M~82 close to the starburst is exposed to a very powerful photoionizing radiation field. A starburst with SFR~$\sim12$ \msunperyr\ generates $Q_{\rm ion}\sim10^{54}$\,s$^{-1}$ ionizing photons according to Starburst99 \citep{Leitherer1999}.
Only a fraction of this radiation will escape the central regions, due to absorption by neutral gas or extinction by dust. It is hard to know the precise escape fraction, $f_{\rm esc}$, but ionizing radiation likely escapes from the central ionized bubble along low density, fully ionized channels. The ratio of Paschen $\alpha$ in the extended extra-planar emission to that centrally concentrated in the starburst suggests $f_{\rm esc}\sim 0.25$, but that is likely an over estimate since the starburst region is heavily extinguished even at 1.875 $\mu$m. Assuming that 90\% of the photons are lost to intervening extinction or neutral material (i.e., $f_{\rm esc}=0.1$) the ionizing photon flux in a region 200~pc (the half-way travel distance in our mosaic discussed above, and coincidentally also the approximate radius of the starburst region) away from a compact starburst will be $F_{\rm ion}\sim2\times10^{10}\,(f_{\rm esc}/0.1)$\,s$^{-1}$\,cm$^{-2}$ (comparable to the ionizing radiation field a few pc away from an O7 star). 

\rev{For neutral material to survive while immersed in such a radiation field, it must have a minimum density.}
For a region with a size of $L\approx0.3\arcsec-0.5\arcsec$ ($5-9$~pc), the typical width of many of the filamentary features, balancing available photon flux against recombination requires a ``threshold'' density of at least $n_{\rm thr}\approx\sqrt{F_{\rm ion}/(\alpha_B L)}\sim50 \sqrt{f_{\rm esc}/0.1}$\,\percmcu\ to stay neutral ($\alpha_B\sim4\times10^{-13}$\,cm$^3$\,s$^{-1}$ is the recombination rate of H).
Radio recombination line measurements \rev{over large scales} point to a high density ionized gas component, which could be in part newly ionized gas \citep{Seaquist1996}. 
Given the ionizing flux computed above, the photoionization timescale is $\tau_{\rm ion}=(F_{\rm ion}\,\sigma_{\rm ion})^{-1}\sim0.3\,(f_{\rm esc}/0.1)^{-1}$~yr for neutral gas (where $\sigma_{\rm ion}$ is the photoionization cross-section of H), and the recombination timescale is \rev{$\tau_{\rm rec}=(\alpha_B\,n)^{-1}\lesssim1,600\,(50\,\percmcu/n)$~yr for ionized gas}. Since these timescales are short compared to the time to traverse a few parsecs at the flow velocity, it is likely that ionization equilibrium is satisfied, although there may be large local variations in ionization state due to lines of sight strongly attenuated by patchy overdense gas. On average, we expect most neutral gas to have densities larger than $n_{\rm thr}$ or be otherwise quickly ionized, \rev{although this will be strongly dependent on $f_{\rm esc}$ and distance to ionizing sources.}


Neutral gas that survives the gauntlet (by having high enough volume and column density to self-shield) and moves away from the central source is likely to \rev{stay neutral} for a long time \citep[the timescale to thermal conduction evaporation is long, $\tau_{\rm evap}\gtrsim10$~Myr depending on density and mass,][]{Micelotta2010b}, and pockets of dense, cool gas may even act as condensation seeds and accrete from their hot surroundings \citep{Gronke2018,Gronke2020}. Measurements in the outflow find column densities of neutral atomic gas $N_{\rm H}\sim3-10\times10^{20}$~\percmsq\ (and similar values for molecular gas) on angular scales of 20\arcsec\ close to M~82 \citep{Leroy2015b,Martini2018}. This neutral gas component of the outflowing material is highly clumped on sub-resolution scales, as suggested by the large scale CO interferometer mapping \citep{Krieger2021}, reaching much higher column density than measured above.  

The picture that emerges from these considerations is that the observed PAH emission likely comes from a mix of neutral and photoionized material. Part of the PAHs are associated with (and protected by) fairly dense and thus likely molecular gas. Not all of this molecular gas may emit in CO, since protecting CO molecules requires higher column densities than it does for H$_2$ in order to build up shielding to the photodissociating UV radiation \citep[see, e.g.,][]{Bolatto2013}. Also, some of this \rev{molecular} gas may be below the sensitivity limit of the existing CO interferometric observations \citep{Krieger2021}. Mid-IR spectroscopy of \htwo\ transitions would be necessary to definitively establish the correlation between 3.3 $\mu$m and molecular emission on small scales.

Because of the morphological similarity between the PAH and ionized gas images, we suggest that another part of the PAH emission arises from photoionized gas, which is easily maintained given the high luminosity of the starburst. Since the intensity of recombination and free-free emission depend sharply on the gas density, bright emission indicates large ionized gas densities, likely occurring at the interface between neutral material \rev{at the photodissociation region} and ionized material \rev{in its vicinity. Projection effects are likely important.} Part of the PAH-emitting material may be located at the walls of the outflow cone around the edges of the starburst, where neutral and molecular gas are advected into the flow, and seen in projection against the central regions of the outflow. 

\section{Conclusions}
\label{sec:conclusions}

We present new, very high resolution infrared images obtained by JWST of the central starburst region and the base of the galactic outflow in M~82 (Figures \ref{fig:3color1}, \ref{fig:3color2}). These images reveal a highly textured network of filamentary and bubble-like structures in the dust component traced by 3.35 $\mu$m PAH emission (Figures \ref{fig:pah}, \ref{fig:unsharp}), as the material is entrained into the wind at the base of the galactic outflow present in M~82. The observed filaments have thickness of $5-9$~pc, approaching the JWST resolution of $\sim2$~pc, lengths that can reach $\sim100$~pc, and associations of these filaments form prominent plumes of emission. The bubble-like structures also reach sizes of $\sim100$~pc. We suggest the latter arise from super-bubbles pumped by individual SNe events ejecting material, but it is also possible they are due to wind instabilities, for example due to cosmic rays (\S\ref{sec:structure}).  \rev{The filaments of parsec-scale width observed in PAH emission (Figure \ref{fig:unsharp}) are likely related to ablation and shredding of dense clumps of material by the galactic wind \citep{Schneider2020,Abruzzo2022,Fielding2022}, possibly located in the walls of the ionized outflow, although the precise mechanisms leading to their structure remain to be determined.}

The structure of the PAH emission resembles closely that of the ionized gas, revealed by recombination in Paschen $\alpha$ and free-free emission at 6 GHz, and has some correspondence with the CO emission (Figure \ref{fig:tracers}). The 6 GHz to Paschen $\alpha$ ratio in the ionized gas filaments indicates that their 6 GHz emission is thermal, and hence they do not appear likely to be due to very strong magnetic fields. We discuss the conditions in the material associated with the PAH emission, finding that the most likely scenario is a combination of PAHs embedded in neutral/molecular gas and photoionized gas, \rev{likely associated with photodissociation regions}. Companion papers explore the quantitative correlations between PAHs and other gas tracers (Fisher et al., in prep.) and the statistics and properties of massive star clusters in the starburst (Levy et al., in prep.). 

\begin{acknowledgments}
\rev{We acknowledge the comments from Bruce Draine, Brandon Hensley, and the anonymous referee, which helped improve this manuscript.} This work is based on observations made with the NASA/ESA/CSA James Webb Space Telescope. The data were obtained from the Mikulski Archive for Space Telescopes at the Space Telescope Science Institute, which is operated by the Association of Universities for Research in Astronomy, Inc., under NASA contract NAS 5-03127 for JWST. These observations are associated with program JWST-GO-01701. Support for program JWST-GO-01701 is provided by NASA through a grant from the Space Telescope Science Institute, which is operated by the Association of Universities for Research in Astronomy, Inc., under NASA contract NAS 5-03127. 

We gratefully acknowledge Alyssa Pagan at the Space Telescope Science Institute for her beautiful multicolor rendition of the data used in Figures \ref{fig:3color1} and \ref{fig:3color2}. A.D.B. and S.A.C. acknowledge support from the NSF under award AST-2108140. R.C.L. acknowledges support for this work provided by a NSF Astronomy and Astrophysics Postdoctoral Fellowship under award AST-2102625. 

R.H.-C. thanks the Max Planck Society for support under the Partner Group project ``The Baryon Cycle in Galaxies'' between the Max Planck Institute for Extraterrestrial Physics and the Universidad de Concepci\'on. R.H-C. also gratefully acknowledge financial support from ANID BASAL project FB210003. 

R.S.K. and S.C.O.G. acknowledge funding from the European Research Council via the Synergy Grant ``ECOGAL'' (project ID 855130), from the German Excellence Strategy via the Heidelberg Cluster of Excellence (EXC 2181 - 390900948) ``STRUCTURES'', and from the German Ministry for Economic Affairs and Climate Action in project ``MAINN'' (funding ID 50OO2206). V. V. acknowledges support from the scholarship ANID-FULBRIGHT BIO 2016 - 56160020, funding from NRAO Student Observing Support (SOS) - SOSPADA-015, and funding from the ALMA-ANID Postdoctoral Fellowship under the award ASTRO21-0062. I.D.L. acknowledges funding support from the European Research Council (ERC) under the European Union’s Horizon 2020 research and innovation programme DustOrigin (ERC-2019-StG-851622) and funding support from the Belgian Science Policy Office (BELSPO) through the PRODEX project “JWST/MIRI Science exploitation” (C4000142239).  L.L. acknowledges that a portion of their research was carried out at the Jet Propulsion Laboratory, California Institute of Technology, under a contract with the National Aeronautics and Space Administration (80NM0018D0004). The work of  E.C.O. was supported in part by grant 510940 from the Simons Foundation.
\end{acknowledgments}

\facilities{JWST:NIRCam, VLA, HST, IRAM:NOEMA, IRAM:30m}
\software{JWST Calibration Pipeline \citep{Bushouse2023}, JHAT \citep{Rest2023}, Matlab}
\bibliographystyle{aasjournal}
\bibliography{references}

\end{document}